\documentclass[twocolumn,showpacs,preprintnumbers,amsmath,amssymb,floatfix,nofootinbib,]{revtex4}
\usepackage{graphicx}
\usepackage[dvips]{epsfig}
\usepackage{dcolumn}
\usepackage{bm}
\usepackage{amsthm}
\usepackage{amsmath}
\usepackage{amsfonts}

\newcommand{\be}{\begin{equation}}
\newcommand{\ee}{\end{equation}}
\newcommand{\ba}{\begin{eqnarray}}
\newcommand{\ea}{\end{eqnarray}}

\newcommand{\n}[1]{\label{#1}}
\newcommand{\eq}[1]{(\ref{#1})}
\newcommand{\pa}{\partial}
\newcommand{\ve}{\varepsilon}

\newcommand{\hh}{\, ,\hspace{0.5cm}}
\newcommand{\hhh}{\, ,\hspace{0.2cm}}

\begin{document}

\title{Brane Holes}
\author{Valeri P. Frolov}
\email{vfrolov@ualberta.ca}
\affiliation{Theoretical Physics Institute, University of Alberta,
Edmonton, AB, Canada,  T6G 2G7}
\author{Shinji Mukohyama}
\email{shinji.mukohyama@ipmu.jp}
\affiliation{
Institute for the Physics and Mathematics of the Universe (IPMU)\\ 
The University of Tokyo, 
5-1-5 Kashiwanoha, Kashiwa, Chiba 277-8582, Japan}
\date{\today}

\begin{abstract}
 The aim of this paper is to demonstrate that in models with large extra
 dimensions under special conditions one can extract information from
 the interior of $4D$ black holes. For this purpose we study an induced
 geometry on a test brane in the background of a higher dimensional
 static black string or a black brane. We show that at the intersection
 surface of the test brane and the bulk black string/brane the induced
 metric has an event horizon, so that the test brane contains a black
 hole. We call it a {\it brane hole}. When the test brane moves with a
 constant velocity $V$ with respect to the bulk black object it also has
 a brane hole, but its gravitational radius $r_e$  is greater than the
 size of the bulk black string/brane $r_0$ by the factor
 $(1-V^2)^{-1}$. We show that bulk `photon' emitted in the region
 between $r_0$ and $r_e$ can meet the test brane again at a point
 outside $r_e$. From the point of view of observers on the test brane
 the events of emission and capture of the bulk `photon' are connected 
 by a spacelike curve in the induced geometry. This shows an example in
 which extra dimensions can be used to extract information from the
 interior of a lower dimensional black object. Instead of the bulk black
 string/brane, one can also consider a bulk geometry without horizon. We
 show that nevertheless the induced geometry on the moving test brane
 can include a brane hole. In such a case the extra dimensions can be
 used to extract information from the complete region of the brane hole
 interior. We discuss thermodynamic properties of brane holes and
 interesting questions which arise when such an extra dimensional
 channel for the information mining exists. 
\end{abstract}

\pacs{04.70.Bw, 04.70.-s, 04.25.-g \hfill
{\bf Alberta-Thy-14-10, IPMU-10-0219}}
\maketitle

\section{Introduction}

Models with large extra dimensions have been `popular' and intensively
discussed since
1998~\cite{ArkaniHamed:1998rs,Antoniadis:1998ig,ArkaniHamed:1998nn}. In  
these models our four dimensional spacetime $\Sigma$ is considered as a
brane embedded in a higher dimensional bulk space. The main purpose of
the present paper is to demonstrate that in such models under special
conditions extra dimensions can be used to extract information from the
interior of a four dimensional black hole. This happens when the four
dimensional surface $\Sigma$ representing our world is not
geodesic. Consider two points on $\Sigma$ and suppose that they can be
connected by a geodesic in the bulk spacetime. Let us assume that there 
also exists a curve on $\Sigma$ which connects the two points and which
is geodesic in the induced geometry. In this case the geodesic distance
between the two points along the bulk geodesic is in general different
from the geodesic distance in the induced geometry. Under special
conditions, for two points on $\Sigma$ separated by a spacelike induced
interval, there may exist causal curves connecting them through the bulk
spacetime. An observer on the brane $\Sigma$ would describe this
situation by saying that the extra dimensions provide one with a channel
of information exchange with an effective super-luminal velocity. 

A simple model demonstrating such a possibility was considered in
\cite{Frolov:1995vp}. In the paper a stationary cosmic string in the
Kerr geometry was studied. The $2D$ induced metric on the string
worldsheet has a horizon at its intersection with the ergosurface of the
bulk geometry. It was shown that the $2D$ black hole produces the
Hawking radiation of the string transverse degrees of freedom
\cite{Frolov:1995qp}, so that the cosmic string can be used to mine
energy from the bulk black hole \cite{Frolov:2000kx}. 

In this paper we study more `realistic' case where a brane representing
our $4D$ world is embedded in a higher dimensional bulk spacetime. We 
neglect effects connected with the thickness of the brane. We also
neglect the gravitational field generated by the brane and use the test
brane approximation. To simplify the presentation in the most of the
paper we discuss the case in which the bulk space has five
dimensions. Generalization to other dimensions of the brane and the bulk 
spacetime is straightforward and is briefly discussed at the end of the
paper. We discuss two models. In the first model the test brane is
moving with a constant velocity (as measured by a distant observer) in
the bulk space with a black  string. We show that if $r_0$ is the
gravitational radius of the black string, the induced geometry has a
$4D$ black hole with the larger radius $r_e=(1-V^2)^{-1}r_0$. We
demonstrate that the test brane embedding is not geodesic, and extra
dimensions can be used to extract information from the region between
$r_0$ and $r_e$ of the induced brane hole. In the second model we
replace a bulk black string by a spacetime with a static massive thin
shell of mass $M$ and radius $r_s$. We shall see later that the
gravitational radius of the shell is $r_0=2M-M^2/r_s\leq r_s$ and that
the bulk spacetime is regular and does not contain a bulk black
object. Nonetheless, if $r_e$ is greater than $r_s$ then the induced
metric has a brane hole. We demonstrate that in this model the complete
brane hole interior is `visible' through extra dimensions.  

The existence of the extra dimensional `window' for observing'  the
brane hole interior raises a number of interesting questions. Some of
them will be briefly discussed in the paper. 

The rest of this paper is organized as follows. In
Sec.~\ref{sec:bs-ds} two five-dimensional geometries describing a black
string and a dark shell at rest are presented. In Sec.~\ref{sec:boost}
these static bulk geometries are boosted so that they describe
a moving string and a moving dark shell, respectively. 
Sec.~\ref{sec:testbrane} describes a test brane in the background of the
boosted black string and shows that the induced metric on the test brane
contains a brane hole. Sec.~\ref{sec:braneholes} shows some
thermodynamic properties of brane holes. In
Sec.~\ref{sec:informationmining} it is shown that information can be
tranmitted via the bulk space from the inside of a brane hole to the
outside. Sec.~\ref{sec:darkshell} describes a test brane in the boosted
dark shell geometry and shows that the induced metric can still have a
brane hole. Sec.~\ref{sec:summary} is devoted to summary of the paper
and discussions. 

\section{Black string and dark shell}
\label{sec:bs-ds}

In this paper we study properties of a moving test brane in an external
gravitational field. For the latter we use two models: (i) a bulk black
brane, and (ii) a dark shell model. In this subsection we discuss the
first model. For simplicity we assume the number of dimensions of the
bulk spacetime is equal to five, so the bulk black brane in fact is a
black string. We also assume that the test brane has a codimension one,
so that the induced geometry on it is four dimensional. In the both
cases one can compactify the metric in the fifth flat dimension, so that
the corresponding $z$-coordinate becomes periodic with some period
$L$. Such a compactification as well as generalizations to higher number
of dimensions, which is quite straightforward, are briefly discussed in
Appendices. 

\subsection{Bulk black string}

The $5$-dimensional metric of a black string is ($a,b=0,1,\ldots,4$) 
\ba\n{blackstring}
dS^2&=&g_{ab}dy^a dy^b=-\Phi d\tilde{t}^2+{dr^2\over \Phi}
+r^2 d\omega^2 +d\tilde{z}^2\, ,\\
\Phi&=&1-\varphi\hh \varphi=1/r\, ,
\ea
where $d\omega^2$ is a line element of a unit sphere $S^2$, 
\be
d\omega^2=d\theta^2+\sin^2\theta d\phi^2\, .
\ee
This metric is a direct sum of the Schwarzschild metric and a line. We
write the metric in the dimensionless form, that is put the
gravitational radius $2M$ equal to one. To restore proper dimensionality
it is sufficient to multiply coordinates $\tilde{t}$, $\tilde{z}$ and
$r$ by the factor $(2M)^{-1}$ and rescale the metric as 
$dS^2\to (2M)^2 dS^2$. 

Denote
\be
\tilde{\eta}^a=\xi^a_{(\tilde{t})}+v \xi^a_{(\tilde{z})}\, ,
\ee
where $\xi^a_{(\tilde{t})}\pa_a=\pa_{\tilde{t}}$ and
$\xi^a_{(\tilde{z})}\pa_a=\pa_{\tilde{z}}$ are two commuting Killing
vectors generating transformations along $\tilde{t}$ and $\tilde{z}$
coordinates, respectively. The norm of this vector is 
\be
\tilde{\eta}^2=-\Phi+v^2\, .
\ee
The vector $\tilde{\eta}^a$ is timelike for $v^2<\Phi$ and becomes
null when $v=\pm\sqrt{\Phi}$.  This means that there exist a Killing
observer with the velocity $u^a\sim\tilde{\eta}^a$ at the radius $r$
only if its relative velocity $v$ with respect to a rest frame is in the
interval $v\in (-\sqrt{\Phi},\sqrt{\Phi})$. Near the horizon where
$\Phi$ vanishes, this interval shrinks to zero. One can say that the
particle motion in $z$-direction is frozen. 

\subsection{Dark shell}

In our second model we assume that outside some radius
$r_s=1+\varepsilon$  with  small positive $\varepsilon$ the metric
coincides with \eq{blackstring} and inside this radius it is flat 
\be\n{dsm}
dS_-^2=-{\ve \over 1+\ve} d\tilde{t}^2+dr^2+r^2 d\omega^2 +d\tilde{z}^2\,.
\ee
In what follows we denote the external metric as $g_{+ ab}$. We choose
the form of the metric so that all the metric coefficients, except
$g_{rr}$ are continues at the junction surface $\Sigma_{\ve}$. The jump 
of the coefficient $g_{rr}$ implies the jump of an extrinsic
curvature. Thus the spacetime contains a thin massive shell. 

Using Israel's method~\cite{Israel:1966rt} it is possible to find the
shell parameters: the surface energy density $\sigma$ and components of
the pressure $P_{z,\theta}$. 
\ba
8\pi\sigma &= & \frac{2}{1+\ve}
\left(1-\sqrt{\frac{\ve}{1+\ve}}\right), \nonumber\\
8\pi P_z+8\pi\sigma 
&= & 
8\pi P_{\theta}+4\pi\sigma
=
\frac{1}{2\ve^{1/2}(1+\ve)^{3/2}}. 
\ea
Note that the mass of the shell is
\be
 M = 4\pi\sigma r_s^2
 = r_s\left(1 - \sqrt{1-\frac{r_0}{r_s}}\right),
\ee
where $r_0=1$ and $r_s=(1+\ve)$ are the horizon radius of the bulk
black string and the radius of the shell, respectively. Thus we obtain
\be
 r_0 = 2M - \frac{M^2}{r_s}. 
\ee

The red-shift factor on the outer surface of the shell is 
$\sqrt{\ve/(1+\ve)}$. For small $\ve$ this factor is small. This explains
an adopted in this paper terminology a {\em dark shell}. 

\section{Boosted metrics}
\label{sec:boost}

Our purpose is to consider a moving test brane in the described
geometries. We assume that the test brane is asymptotically flat and
moved with a constant velocity $V$ in $z$-direction with respect to
either a black string or a dark shell. This problem is equivalent to the
case when the test brane is asymptotically flat and is at rest, while
the black string and the dark shell moves with the constant velocity
$-V$ in $z$-direction. The gravitational field of such objects can be
easily obtained by a boost transformation. 

\subsection{Boosted black string}

In order to obtain a  metric of a boosted black string, let us make the
following transformation 
\ba\n{boost}
\tilde{t}&=&c t+s z\hhh
\tilde{z}=s t+c z\, ,\\
c&=&\cosh \alpha\hhh
s=\sinh \alpha\, .
\ea
Here $\alpha$ is a boost parameter. This transformation generates the
motion in $z$-direction with the velocity $V=s/c$, and $c$ is the
corresponding Lorentz $\gamma$-factor. Applying this transformation to
the black string metric one obtains 
\be\n{bbulk}
dS^2=-dt^2+dz^2+\varphi (c\, dt+s\, dz)^2+{dr^2\over 1-\varphi}+r^2 d\omega^2\, .
\ee

To compactify the boosted black string metric one need to assume  that
$z-$coordinate  is periodic with the period $L$. Notice that operations
of the boost and compactification do not commute. Thus the geometry of
compactified spacetime of the moving black string is different from the
unboosted one, where the periodicity is imposed in the $\tilde{z}$
coordinate.

The metric \eq{bbulk} has a coordinate singularity at the black string
horizon $r=1$. It can be removed by the following coordinate
transformation 
\be\n{transf}
t=v-c r_*\hhh
z=y+s \ln \Phi \hhh r_*=\int {dr\over \Phi}=r+\ln(r-1)\, .
\ee
The metric in these coordinates is
\ba\n{bbulkreg}
dS^2 & = & -(1-c^2\varphi)dv^2 + (1+s^2\varphi) dy^2
\nonumber\\
 & & 
 - s^2(1+\varphi)(1-c^2\varphi-s^2\varphi^2) dr^2
\nonumber\\
 & & 
 + 2cs\varphi dvdy
 - 2s\varphi (c^2+s^2\varphi) dydr
\nonumber\\
 & & 
 + 2c(1-s^2\varphi-s^2\varphi^2) drdv.
\ea
The new coordinates $(v,r,y,\theta,\phi)$ cover both exterior and
interior of the black string. These coordinates are similar to the
advance time for the Eddington-Finkelstein coordinates in the
Schwarzschild spacetime. By changing the signs in \eq{transf} one can
define $u$ coordinate which is an analogue of the retarded time.

\subsection{Ergoregion}

An important new feature of the spacetime \eq{bbulk} is an existence of
the {\it ergoregion}. The metric \eq{bbulk} has two commuting Killing
vectors $\xi^{a}_{(t)}\partial_a=\partial_t$ and
$\xi^{a}_{(z)}\partial_a=\partial_z$. One has 
\be
(\xi_{(t)})^2=g_{tt}=-(1-c^2\varphi)\, .
\ee
Hence the vector $\xi^{a}_{(t)}$ is timelike at the spatial infinity, it
becomes null at $r=r_{e}=c^2$, and it is spacelike inside this
surface. The infinite red-shift surface $r_{e}$ is located outside the
boosted bulk brane horizon $r=1$. We call the spacetime region between
$r=r_{e}$ and $1$ an {\it ergoregion}, and its external boundary,
$r=r_{e}$ an {\it ergosurface}. 

The ergoregion has an important characteristic property: causal
propagation inside it is always a motion with the decrease of the
coordinate $z$. To demonstrate this let us consider a linear combination
of the Killing vectors 
\be
\eta=\xi^{a}_{(t)}+v\xi^{a}_{(z)}\, .
\ee
Its square is
\be
\eta^2=-(1-c^2\varphi)+2vsc\varphi+v^2(1+s^2\varphi)\, .
\ee
The velocity of an observer who is at rest with respect to the infinity
is proportional to  $\xi^{a}_{(t)}$. An observer with the velocity
directed along the vector  $\eta$ is moving with respect to the rest
frame with the speed $v$. A condition $\eta^2=0$ determines an
intersection of a local null cone with $(t-z)$-plane. Solving this
equation we find 
\be
v_{\pm}={-sc\varphi \pm\sqrt{1-\varphi}\over 1+s^2\varphi}\, .
\ee

Causal motion with constant $r$ is impossible inside the surface
$r=1$. This region is the bulk black string interior. Notice that the
sign of $v_-$ is always negative.  $v_+$ vanishes when 
\be
sc\varphi =\sqrt{1-\varphi}\, .
\ee
This equation has two roots
\be
\varphi=c^{-2}\, \mbox{   and  } \varphi=-s^{-2}\, .
\ee
The second root gives $r=-s^2$ and hence it is unphysical. The first
root determines $r=r_{e}=c^2$. This is an equation of the
ergosurface. Inside the ergosurface both $v_+$ and $v_-$ are
negative. Hence in the ergoregion particles and light always propagate
with decrease of the coordinate $z$. In their motion the radius $r$ can
become smaller or larger. Hence a particle and light can leave the
ergoregion. Inside $1$ the motion always reduces $r$, so that the
horizon $1$ is a surface of `no return'. These properties are similar to 
the properties of stationary rotating black holes. 

\subsection{Boosted dark shell}

The metric \eq{bbulk} describes also the external metric for the boosted
dark shell. The inner metric can be obtained easily by applying the
boost transformation \eq{boost} to \eq{dsm}  and it is of the form 
\be\n{bds}
dS_-^{2}=-{\ve -s^2\over 1+\ve}dt^2+{2sc\over 1+\ve}dt\, dz+{c^2+\ve\over 1+\ve}dz^2+dr^2+r^2 d\omega^2\,.
\ee

\section{Test brane in the boosted black string spacetime}
\label{sec:testbrane}

\subsection{Test brane equation}

We refer to the spacetime with metric \eq{bbulk} as a 
{\it bulk space}. We assume now that in the bulk space there exists a
$4D$ brane which represents our `physical' spacetime. Such a brane is a
$4D$ submanifold $\Sigma$ embedded in the $5D$ bulk manifold. We assume
that the brane `respects' the symmetries of the bulk space, that is it
is static and spherically symmetric and choose the equation for the
embedding  in the form 
\be
F=z-Z(r)=0\,.
\ee
The induced metric on the test brane is ($\mu,\nu=0,1,2,3$)
\ba\n{ind}
ds^2&=&h_{\mu\nu}dx^{\mu}dx^{\nu}=-(1-c^2\varphi)dt^2+ 2sc Z'\varphi dt dr \nonumber\\
&+&\left[ (1+s^2\varphi){Z'}^2+{1\over 1-\varphi}\right]dr^2+r^2 d\omega^2\, .
\ea
Here $(\ldots)'=d(\ldots)/dr$.
The induced metric can be diagonalized  by the following coordinate
redefinition 
\be
T=t-sc\int {dr \varphi Z'\over 1-c^2\varphi}\, .
\ee
In the  new coordinates the metric \eq{ind} is
\ba\n{meT}
ds^2&=&-(1-c^2\varphi)dT^2\nonumber\\
&+&\left[{Z'}^2 {1-\varphi\over 1-c^2\varphi}+{1\over 1-\varphi}\right]dr^2+r^2d\omega\, .
\ea

Simple calculations give the the following expression for the
determinant $g$ of the induced metric 
\be
g= - r^4 \sin^2\theta {{Z'}^2(1-\varphi)^2+1-c^2\varphi\over 1-\varphi}\, .
\ee
We choose the brane action in the form
\ba\n{action}
W&=&\int dT\int_0^{2\pi}d\phi \int_0^{\pi} d\theta \sin\theta\int dr \sqrt{-g}\nonumber\\
&=&4\pi \Delta T\int dr L\\
L&=&r^2\sqrt{ {{Z'}^2(1-\varphi)^2+1-c^2\varphi\over 1-\varphi}}\, .
\ea
Here $\Delta T$ is an interval of the time $T$. An extremum of the
action $W$ determines a minimal surface $\Sigma$, which is the
world-volume of the brane. By varying \eq{action} one obtains the
following equation 
\be\n{tbe}
{d\over dr}\left[{r^2(1-\varphi)^{3/2} Z'\over \sqrt{{Z'}^2(1-\varphi)^2+1-c^2\varphi}}\right]=0\, .
\ee
This equation implies that the expression in the square brackets is a
constant. We denote  this constant by $B$, then one has 
\be\n{zp}
Z'=\frac{B}{1-\varphi}
\sqrt{\frac{1-c^2\varphi}{r^4(1-\varphi)-B^2}}\,.
\ee

\subsection{Induced geometry}

Substitution of \eq{zp} in \eq{meT} gives
\ba\n{meT1}
ds^2&=&-(1-c^2\varphi)dT^2\nonumber\\
&+&\left[{r^4\over r^4(1-\varphi)-B^2}\right]dr^2+r^2d\omega\, .
\ea

The equation (\ref{zp}) contains an arbitrary parameter $B$. It can be
fixed by imposing the condition that the induced metric is regular at
the surface $r_e$ where it crosses the ergosurface. 

The Ricci scalar for the induced metric is
\be
R={ c^2 s^2 r^4-12 B^2 r^2+20 B^2 c^2 r-9 c^4 B^2\over 2 r^6(r-c^2)^2}\, .
\ee
The Ricci scalar of the induced metric is regular for $r>c^2$. In a
general case it is divergent at 
\be\n{bhh}
r=r_{e}=c^2\, .
\ee
The regularity condition at this point singles out a special value of
the integration constant $B$ 
\be\n{BBB}
B=\pm s c^3\, .
\ee
For this value the Ricci scalar takes the form
\be
R={s^2 c^2(r-r_+)(r-r_-)\over 2r^6}\hh
r_{\pm}=(-1\pm\sqrt{10})c^2\, .
\ee
For this choice of $B$ the induced metric is regular at $r=r_{e}$. It
takes the form 
\be\n{regind}
ds^2=-(1-{c^2\over r})dT^2+{r^4\over (r-c^2)U}dr^2+r^2d\omega\, ,
\ee
where
\be\n{eqU}
U=r^3+s^2 (r^2 +c^2r +c^4)\, .
\ee

This geometry represents a $4$-dimensional black hole induced on the
brane. The expression \eq{bhh} gives the size of its horizon and, for
$V\ne 0$, it is greater than that of the bulk black string. It is easy to
understand the physical reason for this. For $V\ne 0$, i.e. for a moving
brane, a null vector tangent to the brane world-volume always has a
non-vanishing $\tilde{z}$-component from the bulk point of view. As a
result, its $r$-component is smaller than the speed of light. This means
that a null geodesic on the brane is easier to be trapped by gravity of
the black string than a radial null geodesic in the bulk. In particular,
there exist outward null geodesics on the brane which start from points
slightly outside the black string horizon and are still trapped by
gravity of the black string. This explains the physical reason why the
horizon defined by the brane-induced geometry is greater than the black
string horizon in the bulk. We shall give an alternative explanation for
this fact at the end of this section.

The proper distance to the brane-hole horizon is finite. At $r=c^2$ one
has $U=c^4(4c^2-3)$. The proper distance from $r=c^2$ to a nearby point 
$r$ is 
\be
\rho \sim {2c^2\over \sqrt{4c^2-3}}\sqrt{r-c^2}\, .
\ee

For the metric \eq{regind}, not only the Ricci scalar but also other
curvature invariants remain finite. (See, for example the explicit form
of the Ricci tensor presented in the Appendix~\ref{app:Ricci}.) This
regularity follows from the following observation. Denote 
\be
r=c^2+{4c^2-3\over 4c^4}\rho^2\, ,
\ee
then near $r\approx c^2$ one has
\be
ds^2\approx -\kappa^2 \rho^2 dT^2+d\rho^2+c^2 d\omega^2\, .
\ee
This metric has the Rindler form in the $(T,\rho)$ sector. At $\rho=0$
there exists a horizon. The coordinate $\rho$ has the meaning of the
proper distance from the horizon, and 
\be
\kappa={\sqrt{4c^2-3}\over 2 c^3}
\ee
is the surface gravity of the horizon.

For the regular test brane one has
\be\n{ZZZ}
Z'=\pm {sc^3 \sqrt{r}\over (r-1)\sqrt{U}}\,.
\ee
Note that this expression is well-defined except at $r=1$, while the
regime of validity of \eq{zp} is more restrictive for other values of
$B$. At large distance $Z'\sim \pm sc^3/r^{7/2}$. Hence 
\be\n{Zasymptotic}
Z\sim Z_0\mp {2\over 5}{sc^3\over r^{5/2}}\, .
\ee
Function $Z(r)$ rather fast reaches its asymptotic value $Z_0$. At the
black string horizon $U=U(1)=c^6$ and one has 
\be\n{ZPAS}
Z'\sim \pm {s\over r-1}\, ,
\ee
so that
\be
Z\sim \pm s\ln (r-1)\,.
\ee

\begin{figure}[htb]
\begin{center}
\includegraphics[width=7.0cm]{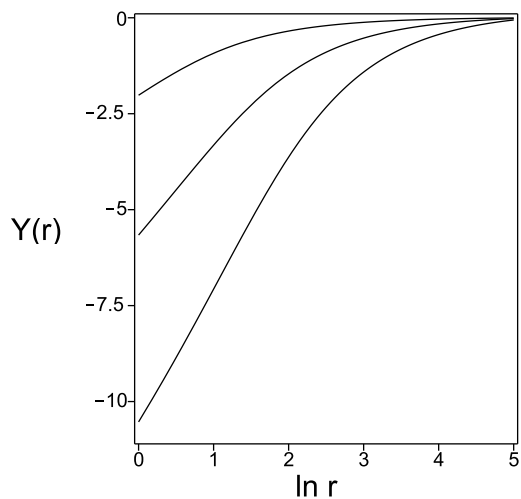}
\caption{The brane is embedded as $z=Z(r)$, or equivalently $y=Y(r)$. In
 this figure, the function $Y(r)$ is shown for $c=1.5, 2.0, 2.5$ (from
 up left to down right) with $Z_0=0$ and the minus sign in
 \eq{Zasymptotic}.} 
 \label{brane} 
\end{center}
\end{figure}

The embedding function $Z(r)$ is singular at the black string horizon
$r=1$. This is an consequence of the coordinate singularity of the
metric \eq{bbulk} at this point. This singularity can be removed by
the coordinate transformations \eq{transf}. In the new coordinates
$(v,r,y,\theta,\phi)$ the brane  equation is $y=Y(r)$ 
\be\n{YYY}
Y'={sc^3 \sqrt{r}\over (r-1)\sqrt{U}}-{s\over r(r-1)}\, .
\ee
We chose the $+$ sign in \eq{ZZZ}. For the opposite sign one needs to
use the retarded time $u$. Relation \eq{ZPAS} shows that $Y(r)$ is 
regular at $r=1$. Figure~\ref{brane} shows some examples of the function
$Y(r)$.

To summarize, the induced metric on the brane is the metric of $4D$
asymptotically flat spacetime with a static black hole in it. We call
this object a {\it brane hole}. The gravitational radius of the brane
hole is $r_{e}=c^2$. And it is located outside the horizon radius of
the bulk black string.

There is a simple alternative explanation why the brane-hole horizon is
at $r=r_{e}=c^2$. The $5D$ mass ${\cal M}$ of the black string is 
\be
{\cal M}=M L\, ,
\ee
where $M$ is a $4D$ mass (in our case $M=1/2$) and $L$ is the length of
the string (or its segment). Thus the 4D mass can be define as 
\be
M={d{\cal M}\over dL}\, .
\ee
The reference frame in which the black string is at rest is a preferable
one. In the reference frame moving with the velocity $V$ the black
string energy is ${\cal E}=\gamma {\cal M}$, where $\gamma$ is the
Lorentz factor, $\gamma=(1-V^2)^{-1/2}$. At the same time because of the 
Lorentz contraction the length element in $z$-direction in the moving
frame is ${\cal L}=\gamma^{-1}L$. As a result, the 4D energy of the
moving black string is 
\be
{\tilde M}=\gamma^2 M\, .
\ee
Now, in our parameterization $c=\gamma$. As a result, the effective
gravitational radius, as measured by an observer moving in
$z$-direction, in our units is $r_e=c^2$. This explains the obtained 
relation \eq{bhh} for the `gravitational radius' of the brane hole. 

\section{Brane holes and their properties}
\label{sec:braneholes}

The surface gravity of the brane hole depends on the test brane velocity 
\be
\kappa={\sqrt{4c^2-3}\over 2 c^3}\, .
\ee
When it is not moving, $c=1$, and the surface gravity $\kappa_0$ 
coincides with the surface gravity of the bulk black string 
\be
\kappa_0=1/2\, .
\ee

\begin{figure}[htb]
\begin{center}
\includegraphics[width=7.0cm]{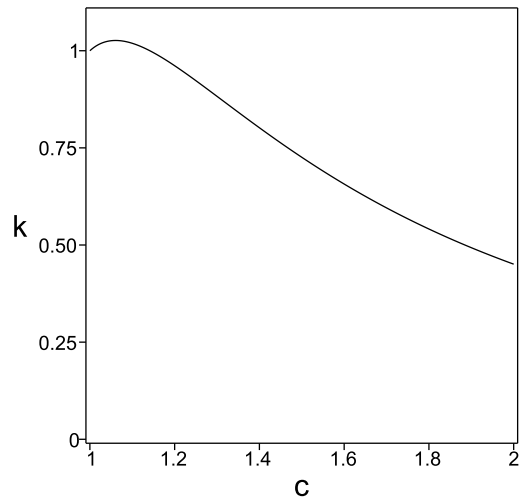}
\caption{A plot of the ratio $k=\kappa/\kappa_0$ as a function of $c$.}
 \label{ratio_kappa} 
\end{center}
\end{figure}

The plot presented in Figure~\ref{ratio_kappa} shows that the ratio
$k=\kappa/\kappa_0$, which is equal to $1$ at $c=1$,  
is greater than one for small non-zero velocity. It reaches the maximum
at $c=3/(2\sqrt{2})$. At  
\be
c={\sqrt{\sqrt{13}-1}\over \sqrt{2}}
\ee
$k$ takes the value $1$ again, and after this it decreases to $0$ when 
$c\to \infty$. 

The brane hole temperature is
\be\n{braneholeT}
\Theta={\kappa\over 2\pi}={\sqrt{4c^2-3}\over 8\pi Mc^3}\,.
\ee
We restored the mass $M$ in this formula. For $c=1$ we obtain the
standard expression for the Hawking temperature of a $4D$ black hole of
mass $M$ 
\be
\Theta_0={1\over 8\pi M}\, .
\ee

As we have seen in the previous section, the horizon radius of the brane
hole is $c^2r_0$, where $r_0=2M$ is the horizon radius of the bulk black
string. This means that the Misner-Sharp mass of the induced metric at
the brane hole horizon is not $M$ but $c^2M$. On the other hand, the ADM
mass, or the Misner-Sharp energy at infinity, is $M$. Thus, we consider
$M$ as energy. 

\begin{figure}[htb]
\begin{center}
\includegraphics[width=7.0cm]{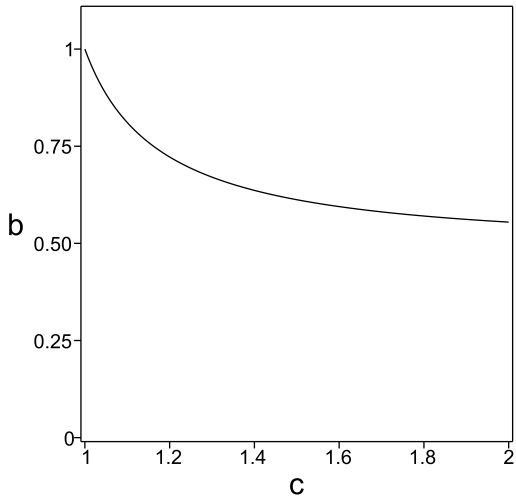}
\caption{A plot of  $\beta$ as a function of $c$.} \label{ratio_S}
\end{center}
\end{figure}

One can easily write the following relation
\be
dM=\Theta dS + \mu_cdc\, .
\ee
Using this first law we obtain the entropy of the brane hole
\be
S={4\pi c^3\over \sqrt{4c^2-3}} M^2 + S_0(c)\, ,
\ee
where $S_0(c)$ is an arbitrary function of $c$. By demanding that $S=0$
for $M=0$, we obtain $S_0(c)=0$ and thus
\be
S={4\pi c^3\over \sqrt{4c^2-3}} M^2\, \hh
\mu_c = \frac{9-8c^2}{2c(4c^2-3)}M\, .
\ee
For zero velocity case the entropy is
\be
S_0=4\pi M^2={1\over 4}{\cal A}_0\, ,
\ee
where ${\cal A}_0$ is the surface area of the $4D$ Schwarzschild black
hole. The surface area of the brane hole is
\be
{\cal A}=16\pi c^2 M^2\, ,
\ee
and one has
\be
S={\beta\over 4}{\cal A}\hh \beta={c\over \sqrt{4c^2-3}}\, .
\ee
As shown in Figure~\ref{ratio_S}, the ratio $\beta$ is $1$ at $c=1$ and
monotonically decreases towards $1/2$. 

\section{Information mining from brane-hole interior through the
 extra-dimensional `window'}
\label{sec:informationmining}

The surface of the test brane besides the induced metric $g_{\mu\nu}$ is
also characterized by the extrinsic curvature $K_{\mu\nu}$ , which
encodes the information about its embedding in the bulk spacetime. The
test brane is a minimal surface and its equation \eq{tbe} is equivalent
to the condition 
\be
Tr K\equiv g^{\mu\nu}K_{\mu\nu}=0\, .
\ee
The calculation of the extrinsic curvature shows that the extrinsic
curvature itself does not vanish (see Appendix~\ref{app:K}), so that the
test brane surface is not geodesic. The latter property means that in a 
generic case a bulk geodesic connecting two point $p$ and $p'$ on the
brane differs from a geodesic connecting these points in the induces
geometry. We demonstrate now that the motion through the bulk can
provide one with a short-cut, so that two points of the brane located at
spacelike interval along the brane can be connected by a causal curve in
the bulk spacetime. 

To illustrate this, let us consider a simple case of a radial null ray
in the bulk space. For this problem it is sufficient to work with a $3D$
metric 
\be 
dS^2=-\Phi d\tilde{t}^2+{dr^2\over \Phi} +d\tilde{z}^2\, .
\ee
The conserved quantities for the radial 
$(\dot{\tilde{t}}, \dot{\tilde{z}},\dot{r})$ motion are 
\be\n{nrcq}
{\cal E}=\Phi \dot{\tilde{t}}\hh K=\dot{\tilde{z}}\, .
\ee
Here a dot denotes derivative with respect to an affine parameter
$\lambda$, ${\cal E}$ is the energy of the photon, and $K$ is the
$z$-component of its momentum. We choose the affine parameter so that it
grows in the `future direction', so that $\dot{\tilde{t}}$ and 
${\cal E}$ are positive. 
For the null ray one has $dS^2=0$. This relation and \eq{nrcq} give 
\be
\dot{r}^2={\cal E}^2-K^2\Phi\hh \dot{\tilde{z}}=K\hh \dot{\tilde{t}}={\cal E}/\Phi\,.
\ee
It is convenient to exclude the affine parameter $\lambda$ and
parameterize the curve by $r$. The corresponding equations are 
\ba
{d\tilde{t}\over dr}&=&{ 1\over \Phi \sqrt{ 1-q^2\Phi}}\, ,\\
{d\tilde{z}\over dr}&=&{ q\over \sqrt{1-q^2\Phi}}\, ,
\ea
where $q=K/{\cal E}$. A solution of this set of first order
equations determine a null ray trajectory
$(\tilde{t}(r),\tilde{z}(r))$. Notice that we chose the sign of the
square root, which enters these expression, to be positive. This
corresponds to photons propagating outwards. 

For points on the moving test brane one has the following relation 
\be
F\equiv s\tilde{t}-c\tilde{z}+Z(r)=0\, .
\ee
The condition that a null ray meets the moving test brane is 
\be
{\cal F}(r)= s\tilde{t}(r)-c\tilde{z}(r)+Z(r)=0\, .
\ee
Suppose this condition is satisfied at some initial point $p$ where
$r=r_1$. In order to determine whether it meets the brane again at
larger value of $r$, it is sufficient to solve the following differential
equation 
\ba
{d{\cal F}\over dr}&=&Q\, ,\\
Q&=&{s-cq\Phi\over \Phi \sqrt{1-q^2\Phi}}+{sc^3\sqrt{r}\over (r-1)\sqrt{U}}\, ,
\ea
with the initial condition ${\cal F}(r_1)=0$. If a solution passes again
through zero at some other radius $r_2$, this will determine another
point of the intersection of the null ray with the test brane.

\begin{figure}[htb]
\begin{center}
\includegraphics[width=7.0cm]{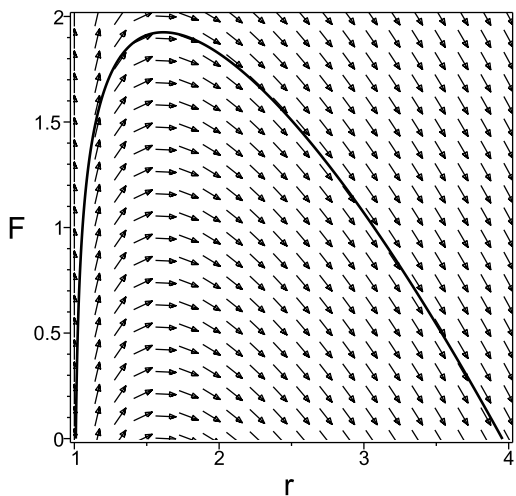}
\caption{An example of a bulk null ray connecting brane hole interior
 with an external region. It shows the function ${\cal F}$ with the
 initial data ${\cal F}=0$ at $r=1.01$ for following set of parameters:
 $s=0.3$, $q=1.0$. This null ray crosses the test brane again near
 $r\approx 4.0$.} \label{f1} 
\end{center}
\end{figure}

Qualitatively it is possible to describe the properties of the function
$Q$ as follows. For small radius $r=1+\ve$, $Q\sim 2s/\ve$ and is
positive. At large radius $r$ it becomes constant and is equal to 
\be
Q\sim {s-c q\over \sqrt{1-q^2}}\, .
\ee
Hence in this region if $q>s/c=V$, the function $Q$ is negative. This
gives us the following asymptotics for the function ${\cal F}$ 
\ba
{\cal F}&\sim &{2s\ve}(r-1-\ve)\hh \mbox{for   } r\approx 1+\ve\, ,\\
{\cal F}&\sim & {\cal F}_0-{q c-s\over \sqrt{1-q^2}} r\, .
\ea
This means that if $q>s/c=V$, the null ray necessarily meets the test
brane again at sufficiently large $r$. The Figure~\ref{f1} shows the
behavior of ${\cal F}$ for a special choice of the parameters. 

\section{Brane holes in the dark shell geometry}
\label{sec:darkshell}

Let us discuss a model of a test brane which is moving with constant
velocity with respect to a dark shell. We again assume that the test
brane is asymptotically flat. Near the shell the brane surface it is
stretched. We focus on the case in which the radius of the shell
$r=1+\ve$ is inside the ergosphere radius $r=c^2$. This gives the
following condition
\be\n{dscond}
\ve <s^2\, .
\ee
We can 'create' and 'destroy' a brane hole by changing the velocity of
the test brane. 

Since outside the dark shell the induced metric $g_{+\mu\nu}$ coincides 
with \eq{regind} it possesses a brane hole horizon. Let us emphasize
that in such a case a brane hole exists even when there is no bulk black
hole. We study now this interesting case in more detail. 

First of all let us calculate the embedding surface of the brane inside
the dark shell and obtain the inner induced metric $ds_-^2$.  The test
brane equation is of the form 
\be
z=Z_-(r)\, .
\ee
Using expression \eq{bds} we find the induced metric
\ba
ds_-^2&=&{1\over 1+\ve}\left[ (s^2-\ve) dt^2+2scZ_-' dt dr \right.\nonumber \\ &+&\left.[1+\ve+(\ve+c^2)(Z_-')^2]dr^2\right]+r^2 d\omega^2\, .
\ea

\begin{figure}[htb]
\begin{center}
\includegraphics[width=7.0cm]{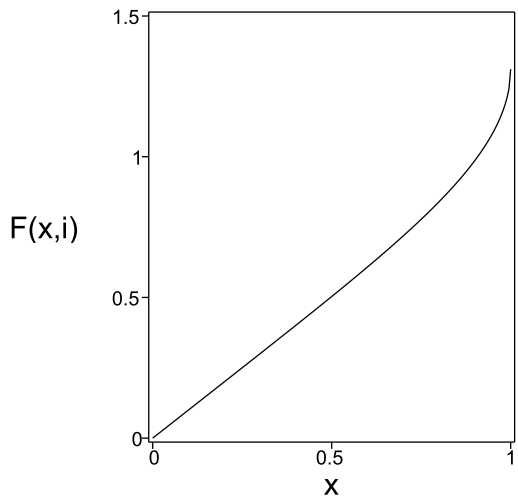}
\caption{A plot of the function $F(x,i)$ .} \label{EllipticF}
\end{center}
\end{figure}

The calculation of the determinant of this metric gives
\be
\sqrt{-\mbox{det}g_-}={\sqrt{\ve (1+(Z_-'))^2-s^2}\over \sqrt{1+\ve}}r^2\sin^2\theta\, .
\ee
Thus the effective action for the test brane is
\ba
W_-&=&{4\pi \Delta T\over \sqrt{1+\ve}}\int dr L_-\, ,\\
L_-&=&r^2\sqrt{\ve (1+(Z_-'))^2-s^2}\, .
\ea
The test brane equation in the inner region
\be
{d\over dr}\left( {\pa L_-\over \pa Z_-'}\right)=0
\ee
has a solution
\be\n{zmb}
Z_-'= \frac{B_-}{\sqrt{\ve}}
\sqrt{\frac{s^2-\ve}{B_-^2-\ve r^4}}\,.
\ee
Here $B_-$ is an arbitrary integration constant. The induced metric for
this solution can be made diagonal by means of the transformation 
\be
t=T-{sc\over s^2-\ve}\int dr Z_-'\, .
\ee
One has
\be\n{inds}
ds_-^2={s^2-\ve\over 1+\ve} dT^2+{1\over \ve-s^2}\left[ \ve(1+(Z_-')^2)-s^2\right] dr^2+r^2 d\omega^2\, .
\ee
The induced metric $ds_+^2$ taken at the shell surface is
\be\n{outds}
ds_+^2={s^2-\ve\over 1+\ve} dT^2+{(1+\ve)^4\over (\ve-s^2)U(r=1+\ve)}dr^2 +r^2 d\omega^2\, .
\ee
Let us notice that the metric coefficient , except $g_{rr}$, in
\eq{inds} and \eq{outds} are continues  on the shell, and
$g_{TT}<0$. From the analysis of the test brane equation it follows that
the following condition must be satisfied on the shell (see
Appendix~\ref{app:junction})  
\be
\frac{Z_-'}{\sqrt{g_{-rr}}} = \frac{Z_+'}{\sqrt{g_{+rr}}}\, .
\ee
Solving this equation one finds the value of $Z_-'$ on the
shell. Equation (\ref{zmb}) at the shell can be used to get
$B_-=\pm sc^3\sqrt{1+\ve}$. Substituting $B_-$ back into \eq{zmb} one
finds the function $Z_-(r)$. The result can be written in the form 
\be
Z_-'={A\over \sqrt{C^2-r^4}}\, ,
\ee
where
\be
A= \pm \frac{sc^3}{\ve}\sqrt{(1+\ve)(s^2-\ve)}
\hh  C=|s|c^3\sqrt{\frac{1+\ve}{\ve}}\, .
\ee
Introducing a new variable $x=r/\sqrt{C}$, one obtains
\be
Z_-=\frac{A}{\sqrt{C}}\int_0^x {dx\over \sqrt{1-x^4}}
=\frac{A}{\sqrt{C}}F(x,i)+Z_{-,0}\, .
\ee
Here $F(x,k)$ is the the incomplete elliptic integral of the first
kind~\footnote{
There are some notational variants for this function. For example, in
\cite{AbramowitzStegun}, the same function is denoted as 
$F(\varphi\backslash\alpha)$ and $F(\varphi |m)$, where $x=\sin\varphi$,
$k=\sin\alpha$ and $m=k^2$.}. 
The constant $Z_{-,0}$ is determined by using the continuity of the
function $Z$ at the shell. The function $F(x,i)$ is shown in the
Figure~\ref{EllipticF}. 

The test brane embedding described in this section is singular at the
origin $r=0$. However, this just indicates breakdown of approximations
which we have implicitly assumed. Thus it is expected that the apparent
singularity at $r=0$ can be resolved if we take into account effects
such as thickness, microscopic physical degrees of freedom of the brane
configuration, etc.   

\section{Summary and discussions}
\label{sec:summary}

We have studied a {\it brane hole}, a black hole induced on a test brane
in the background of a higher dimensional bulk black string/brane. When
the test brane moves with a constant velocity $V$ relative to the bulk
black string/brane, the horizon radius of the brane hole $r_e$ is
greater than that of the bulk black string/brane $r_0$ by the factor
$(1-V^2)^{-1}$. We have shown that bulk  `photons' emitted in the region
between $r_0$ and $r_e$ can meet the test brane again at a point outside
$r_e$. Therefore, a brane hole provides an explicit example in which
extra dimensions can be used to extract information from the interior of
a lower dimensional black object. 

We have also shown that, even if there is no horizon in the higher
dimensional bulk geometry, a moving test brane can still have a brane
hole. As a simple example, we have considered a {\it dark shell}
geometry in which a bulk black string/brane is replaced by a massive
thin shell located outside the would-be horizon. An interesting feature
of this model is that there is no `hidden regions' for the bulk photons,
so that a test brane observer interacting with such photons 
can get information from the complete brane hole interior, including
the central region. 

In order to realize ordinary $4D$ gravity at scales longer than 
$\sim 0.01mm$ in the asymptotic region, the $z$-direction must be
compactified. Also, in the case of a bulk black string/brane, if 
$z$-direction were infinite then the system would be dynamically
unstable due to Gregory-Laflamme instability~\cite{Gregory:1993vy}. It
is well known that compactification of the $z$-direction suppresses the
Gregory-Laflamme instability and can stabilize the bulk black
string/brane. For these reasons, we compactify the $z$-direction by 
imposing a periodic boundary condition. We can perform this
compactification in the rest frame of the test brane. In such a
compactified spacetime the picture of the brane hole and its properties
remain practically unchanged from what we have described. Main
difference which should be mentioned here is that the bulk photons from
the brane hole interior can meet the test brane many times. 

The concept of brane holes opens up new arenas to investigate black hole
evaporation and the information loss problem. Here, let us point out a
couple of interesting possibilities. 

There are physical degrees of freedom on a test brane such as transverse
degrees of freedom of brane fluctuation and matter fields confined on the
brane. If we consider them as free fields and simply quantize them on
a brane hole background then we would conclude that they exhibit Hawking 
radiation with the temperature \eq{braneholeT}. Actually, those degrees 
of freedom inevitably interact with fields propagating in the bulk such
as bulk gravitons. These bulk fields can communicate information stored
inside a brane hole to the outside. Thus, if we take into account
interactions with bulk fields then the dynamics of quantized brane
fields can be quite different. It is certainly worth while seeing if
such interactions can help conveying information stored inside the brane
hole horizon to the outside. 

Another interesting issue would be creation and annihilation of a brane
hole. In the dark shell model, there is a critical velocity $V_c$ below
which the test brane does not contain a brane hole but above which a
brane hole appears on the test brane. Therefore, by controlling the test
brane velocity relative to the rest frame of the dark shell, one can
create and annihilate a brane hole. Possible ways to control the test
brane velocity are accretion and superradiance of Kaluza-Klein
particles. It is interesting to investigate them in details. 

\begin{acknowledgments}
 This work was initiated during V.F.'s visit to IPMU. He is thankful to
 IPMU for their kind hospitality. Part of this work was done during
 APCTP Joint Focus Program: Frontiers of Black Hole Physics. The authors
 thank APCTP for stimulating atmosphere and warm hospitality. One of the 
 authors (V.F.) thanks the Natural Sciences and Engineering Research
 Council of Canada and the Killam Trust for their support. The work of
 S.M. is supported by Grant-in-Aid for Scientific Research 17740134,
 19GS0219, 21111006, 21540278, by World Premier International Research
 Center Initiative (WPI Initiative), and by the Mitsubishi Foundation. 
\end{acknowledgments}

\appendix

\section{Higher dimensional generalizations}

Our starting point is the metric ($a,b=0,1,\ldots,D-1$)
\be\n{blackbrane}
dS^2=g_{ab}dy^a dy^b=-\Phi d\tilde{t}^2+{dr^2\over H}
+r^2 d\omega^2 +d\tilde{z}^2\, .
\ee
The total number of the spacetime dimensions is $D=4+m$,
\be
d\tilde{z}^2 = d\tilde{z}_1^2 + \sum_{i=2}^m dz_i^2
\ee
is the flat $m$-dimensional metric, $d\omega^2$ is a
metric of a unit round $2$-sphere $S^2$. (It is straightforward to
consider an $n$-sphere $S^n$ ($n\geq 3$) instead of a $2$-sphere 
$S^2$. However, in this case the number of non-compact dimensions would 
become more than $4$. Thus, we shall concentrate on the case with
$S^2$.) For 
\be
\Phi=H=1-\varphi\hh \varphi=r_0/r\, ,
\ee
the metric \eq{blackbrane} is a solution of the vacuum Einstein
equation. This solution is a direct sum of  $4D$ Schwarzschild metric
and the $m$-dimensional flat metric. It is called a {\it  black brane}. 
Since we have already considered the case with $m=1$, in this appendix
we shall suppose that $m\geq 2$.

A boosted black brane can be obtained by the transformation
\ba
\tilde{t}&=&c t+s z_1\hhh
\tilde{z}_1=s t+c z_1\, ,\\
c&=&\cosh \alpha\hhh
s=\sinh \alpha\, .
\ea
Let us consider a $(\bar{m}+3)$-brane ($0\leq\bar{m}\leq m-1$) whose 
worldvolume is specified by the embedding 
\be
 z_1 = Z(r), \quad z_j = z_{j,0}\quad (j=\bar{m}+2,\cdots,m),
\ee
where $z_{j,0}$ are constants. This brane fills not only the $4D$
Schwarzschild geometry but also $\bar{m}$-dimensional flat extra
dimensions ($z_2,\cdots,z_{\bar{m}+1}$). One can compactify all $z_i$ 
($i=1,\cdots,m$) on a $m$-torus. 

The rest of the calculations are essentially the same as those presented
in the main text and, thus, we shall not repeat them here. 

\section{Ricci tensor for the induced metric}
\label{app:Ricci}

The Ricci tensor for the metric \eq{regind} is
\ba
R^{\nu}_{\mu}&=&\mbox{diag}(R_0,R_1,R_2,R_3)\, ,\\
R_0&=& 1/4 r^{-6}c^2\left(3c^6-3c^4+2c^4r\right. \nonumber\\
&+&\left. c^2r^2-2c^2 r-r^2\right)\, ,\\
R_1&=&-1/4r^{-6}\left(9c^8-9c^6-6rc^6-5c^4r^2\right.\nonumber\\
&+&\left.6c^4r+5c^2r^2-4c^2r^3+4r^3\right)\, ,\\
R_2&=&-1/2 r^{-6}\left(3c^8-3c^6+rc^6-c^4r\right.\nonumber\\
&+&\left.c^4r^2-c^2r^2+c^2r^3-r^3\right)\, ,\\
R_3&=&-1/2 r^{-6}\left(3c^8-3c^6+rc^6-c^4r\right.\nonumber\\
&+&\left.c^4r^2-c^2r^2+c^2r^3-r^3\right)\, .
\ea

\section{Extrinsic curvature}
\label{app:K}

In the absence of the boost, when $c=1$, the solution for the brane is
$z =0$. The induced metric coincides with the $4D$ Schwarzschild
metric. The radius of the brane hole is $r_0$. The embedding equation
is symmetric with respect to reflection $z\to -z$. As a result, the
brane surface is geodesically embedded. This property is not true any
more when one has a non-vanishing boost. To see this explicitly let us
calculate the extrinsic curvature. 

Using the coordinates $(t,r,\theta,\phi,z)$ in the bulk space, the $5D$
metric is 
\ba
dS^2&=&-(1-c^2/r)dt^2+{2sc\over r} dt dz+{dr^2\over 1-1/r}\nonumber\\
&+&r^2 d\omega^2+(1+s^2/r)dz^2\,,
\ea
and the equation of the brane $\Sigma$ is $F=z-Z(r)=0$.
A unit normal vector to the surface $\Sigma$ is
\be
n_{\mu}=(0,-{s c^3\over r(r-1)}, \sqrt{{U\over r^3}},0,0)\, .
\ee
Here $U$ us defined by \eq{eqU}.

Let us introduce the following vectors
\ba
e_{\hat{0}}^{a}&=&(\sqrt{{r\over r-c^2}},0,0,0,0)\, ,\nonumber\\
e_{\hat{1}}^{a}&=&({s^2 c^4\over \sqrt{r-c^2}r^{3/2}(r-1)},{\sqrt{(r-c^2) U}\over r^2},0,0,{c^3 s \sqrt{r-c^2}\over r^{3/2}(r-1)})\, ,\nonumber\\
e_{\hat{2}}^{a}&=&(0,0,r^{-1},0,0)\, ,\\
e_{\hat{3}}^{a}&=&(0,0,0,(r\sin\theta)^{-1},0)\, .
\ea

The vectors $e_{\hat{\mu}}^{a}$ are tangent to $\Sigma$,
$e_{\hat{0}}^{a}$ being a timelike. The  vectors $e_{\hat{\mu}}^{a}$ and
$n^a$ are mutually orthogonal and have a unit norm. 

The extrinsic curvature is 
\be
K_{\hat{\mu}\hat{\nu}}=e_{\hat{\mu}}^{a}e_{\hat{\nu}}^{a} n_{a|b}. 
\ee
The non-vanishing components of the extrinsic curvature are
\ba
K_{\hat{0}\hat{0}}&=&{s c^5\over 2r^3(r-c^2)}\, ,\\
K_{\hat{1}\hat{1}}&=&{s c^3(4r-3c)\over 2r^3(r-c^2)}\, ,\\
K_{\hat{0}\hat{1}}&=&{s c\over 2r(r-c^2)}\, ,\\
K_{\hat{2}\hat{2}}&=&K_{\hat{3}\hat{3}}=-{s c^3\over r^3}\, .
\ea
Direct check shows that 
\be\n{trk}
\mbox{Tr} K=-K_{\hat{0}\hat{0}}+K_{\hat{1}\hat{1}}
+K_{\hat{2}\hat{2}}+K_{\hat{3}\hat{3}}=0\,.
\ee
The relation $\mbox{Tr} K=0$ must be valid, since the surface $\Sigma$
is a minimal surface. Thus \eq{trk} is simply a consistency check of the
calculation. 

\section{Junction across dark shell}
\label{app:junction}

We consider a $5$-dimensional geometry without horizon
\be
dS^2 = -dt^2 + dz^2 + \tilde{\varphi}(r)(cdt+sdz)^2
+ \frac{dr^2}{H(r)} + r^2d\omega^2, 
\ee
where $c=\cosh\alpha$, $s=\sinh\alpha$ and $\alpha$ is a boost
parameter. For a brane embedding of the form
\be
z = Z(r),
\ee
the induced metric is
\be
ds^2 = -(1-c^2\tilde{\varphi})dT^2
+ \left[ \frac{1-\tilde{\varphi}}{1-c^2\tilde{\varphi}}(Z')^2 
+\frac{1}{H}\right]dr^2 + r^2 d\omega^2,
\ee
where
\be
 T = t - sc\int\frac{dr\tilde{\varphi}Z'}{1-c^2\tilde{\varphi}}.
\ee
Thus the brane action is
\be
 W = \int dT\int_0^{2\pi}d\phi\int_0^{\pi}d\theta\int dr
  \sqrt{-g} = 4\pi \Delta T \int dr L,
\ee
where
\be
 L = r^2
  \sqrt{\frac{(1-\tilde{\varphi})H(Z')^2+1-c^2\tilde{\varphi}}{H}}. 
\ee
The equation of motion is
\be
 B' = 0, \quad 
  B \equiv \frac{r^2(1-\tilde{\varphi})H^{1/2}Z'}
  {\left[(1-\tilde{\varphi})H(Z')^2+1-c^2\tilde{\varphi}\right]^{1/2}}.
\ee
Thus, $B$ is constant. The definition of $B$ can be solved with respect
to $Z'$ as
\be
 Z' = B\sqrt{\frac{1-c^2\tilde{\varphi}}
  {(1-\tilde{\varphi})H[r^4(1-\tilde{\varphi})-B^2]}}
\ee

Let us now introduce two small parameters $\ve$ and
$\tilde{\ve}$ so that
\be
 0 < \tilde{\ve} \ll \ve < 1,
\ee
and smoothly connect the flat spacetime to a curved spacetime as
\ba
\tilde{\varphi}(r) & = & 
 \left\{
  \begin{array}{cl}
   \frac{1}{1+\ve} & (r < 1+\ve-\tilde{\ve}/2)\\
   \tilde{\varphi}_*(r) & 
    (1+\ve-\tilde{\ve}/2\leq r\leq 1+\ve+\tilde{\ve}/2)
    \\
   \varphi(r) & 
    (r>1+\ve+\tilde{\ve}/2)
  \end{array}
	       \right.,\nonumber\\
 H(r) & = & 
 \left\{
  \begin{array}{cl}
   1 & (r < 1+\ve-\tilde{\ve}/2)\\
   H_*(r) & 
    (1+\ve-\tilde{\ve}/2\leq r\leq 1+\ve+\tilde{\ve}/2)
    \\
   1-\varphi(r) & 
    (r>1+\ve+\tilde{\ve}/2)
  \end{array}
	       \right. ,\nonumber\\
\ea
where $\varphi(r)=1/r$ and the subscript ``*'' represents functions
smoothly connecting the inside region and the outside region. 

For $s^2>\ve$, the induced metric on the brane has a horizon. The
regularity of the horizon determines the constant $B$ as
\be
 B = \pm sc^3.
\ee
Thus, we obtain
\be
 Z' = 
 \left\{
  \begin{array}{cl}
   \pm sc^3\sqrt{\frac{(1+\ve)(s^2-\ve)}
    {\ve[(1+\ve)s^2c^6-\ve r^4]}}
    & (r < 1+\ve-\tilde{\ve}/2)\\
   \pm\frac{sc^3\sqrt{r}}{(r-1)\sqrt{U}}
& 
    (r>1+\ve+\tilde{\ve}/2)
  \end{array}
	       \right. ,
\ee
where
\be
 U = r^3 + s^2(r^2+c^2r+c^4). 
\ee
Finally, we can safely take the limit $\tilde{\ve}\to 0$. We
conclude that $\sqrt{H}Z'$ is continuous in this limit.

\end{document}